\documentclass{elsart}
 \usepackage{amsmath,bm,graphicx,enumerate,array}

 \newcommand{\Half}{\textstyle\frac{1}{2}}
 
 \newcommand{\D}{\textstyle{\rm d}}
 \newcommand{\E}{\textstyle{\rm e}}

\begin{document}

 \begin{frontmatter}

 \title{Quarkonium bound-state problem in momentum space revisited}

 \author{A. Deloff}
 \address{Institute for Nuclear Studies, Warsaw, Poland}

 \begin{abstract}
 A semi-spectral Chebyshev 
 method for solving numerically singular integral
 equations is presented and applied in the quarkonium bound-state
 problem in momentum space. The integrals containing both,
 logarithmic and Cauchy singular kernels, can be evaluated 
 without subtractions by dedicated automatic quadratures.
 By introducing a Chebyshev mesh and using the Nystrom algorithm
 the singular integral equation is converted into an algebraic
 eigenvalue problem that can be solved by standard methods.
 The proposed scheme is very simple to use, is easy in programming and highly accurate.
 \end{abstract}

 
\end{frontmatter}
%%%%%%%%%%%%%%%%%%%%%%%%%%%%%%%%%%%%%%%%%%%%%%%%%%%%%%%%%%%%%%%%%%%%%%%%%
 \section{Introduction}
In  a recent work \cite{deloff} we have advocated  
the Chebyshev semi-spectral method demonstrating  its efficiency in
solving some typical differential and integral equations emerging in quantum mechanics. 
The present paper is in the same vein but  here we wish 
to focus our attention solely on  the heavy quarkonium momentum space  bound-state problem.
Admittedly, the problem is not new but our incentive here is  to examine
the effectiveness of the semi-spectral approach in solving strongly
singular integral equations. Since the latter topic was beyond the scope of \cite{deloff},
this work may be regarded as an immediate continuation of  the previous paper. 
\par
 We would like to believe that the presented method will be useful also outside 
 quantum mechanics,  especially that
 strongly singular integral equations are encountered in many areas of science and engineering. 
 The well known physical applications comprise the quantum mechanical
  scattering problem,  the Omnes formulation \cite{omnes} of the final-state-interaction,
 radiative transfer, neutron transport \cite{chandra} etc. 
 The list of engineering applications is by no means restricted to
 the widely known aerofoil problem \cite{wing} and, indeed,
 many important problems of engineering mechanics like elasticity, plasticity,
 fracture mechanics, etc. may be also efficiently expressed in terms of singular 
 and hypersingular integral equations. 
 Because it is not always possible to find explicit solutions to the problems posed,
 much attention has been devoted to approximate methods. It is interesting to note that
 even when an analytic solution is known, quite often the latter takes 
  the form of a singular integral whose numerical evaluation
 might be more complicated than a numerical solution of the integral equation. 
\par
 A hypersingular integral equation arises in quantum mechanics 
 already at a quite elementary level when the 
 linear potential bound-state problem, easily tackled in configuration space, 
  is approached in momentum space. 
 This problem is far from academic since the linear potential plays an important role
 not only in atomic physics where it is associated with the hydrogen radial Stark 
 effect but also in particle
 physics serving as a simple confinement model of QCD.
 Although, in principle,
 QCD alone should describe the spectroscopy of heavy quarkonia but 
  the implementation of such program is very difficult and instead
 various phenomenological models incorporating  some QCD  properties have been developed 
(for a recent review of quarkonium physics  and references to the literature cf. \cite{cern}).
The QCD motivated quark potential models have played
 a prominent role in understanding quarkonium spectroscopy  and 
are capable of reproducing with surprising accuracy 
a sizable part of the meson and baryon properties. 
The non-relativistic potential  approach may be justified by the fact that
 the bottom quark and, perhaps  to a
lesser extent, also the charmed quark have
 masses that are large in comparison with $\Lambda\;$-- 
 the typical QCD hadronic mass  scale. 
The quark--antiquark potential has been tailored to mock up the properties expected from QCD 
and  the different potential shapes set up in the early days after  years of research have
 evolved to a common form that one might expect from the asymptotic 
limits of QCD. The prototype for these
potentials is still the popular Cornell potential \cite{cornell} 
including the one-gluon-exchange Coulomb potential
supplemented by  a  linear potential simulating  confinement,  as expected from QCD.
Therefore, this potential will be also considered in this paper.  
\par
 Obviously, the non-relativistic potential model can not be pushed
beyond certain  limits and for systems containing one light quark
a complete disregard of relativistic effects might be a serious omission.
In addition to that, it was somewhat embarrassing 
when people realized \cite{non-rel} that within the non-relativistic formalism  
the mesons containing a light quark  might be more massive than a
meson composed with heavier quarks. 
These difficulties could be aleviated at the expense of a semirelativistic
treatment where the relativistic expression for the energy is used.
A popular relativistic extension of the Schr\"{o}dinger equation
is the spinless Salpeter equation
\begin{equation}
\label{I1}
\left[\sqrt{\bm{p}^2+m_1^2}+\sqrt{\bm{p}^2+m_2^2} + V(r)\right]\,\Psi(\bm{r}) = E\,\Psi(\bm{r})
\end{equation}
where $m_1,\, m_2$ are the quark masses, $\bm{p}$ is the c.m. momentum, 
$V(r)$ denotes the quark-antiquark potential and $E$ is the eigenenergy.
Since in such case the Laplacian operator appears under a 
square root,   the coordinate space is rather unwieldy
for solving the bound state problem  and the 
momentum space seems to be the most natural alternative.  
 Indeed, in momentum space  the energy operator is diagonal and
the difference in computational effort between non-relativistic and semi-relativistic
treatment is minor.  Although the momentum space approach solves some problems
automatically but at the same time it does create another difficulty 
 in that the  quark-antiquark potential 
gives rise to a singular kernel in the appropriate integral equation.  
Whilst the Coulomb potential 
yields a kernel  with a
logarithmic singularity that can be removed by subtraction
\cite{lande}, the kernel
 associated with a linear potential exhibits a double-pole singularity
for which the subtraction scheme is insufficient. 
To clarify this important point  let us consider  just the linear potential
 for simplicity restricting our attention to a zero orbital momentum state.
 The  potential  term  that enters the appropriate wave equation involves
the integral with a double pole singularity 
\begin{equation}
\int_0^\infty\dfrac{k^2\,\phi(k)\,\D k}{(k^2-p^2)^2}=\displaystyle
\int_0^\infty\left\{k^2\,\dfrac{\phi(k)-\phi(p)}{k^2-p^2}-\Half p\, \phi^{\,\prime}(p)\right\}
\dfrac{\D k}{k^2-p^2},
\label{I0}
\end{equation}  
where $\phi(k)$ is the wave function, $\phi^{\,\prime}(p)$     denotes the derivative
and $p$ is a real parameter.  It may be easily verified that
the two extra terms occurring on the right hand side  of (\ref{I0})
can be supplemented with impunity because
the integrals multiplying, respectively, 
 $\phi(p)$ and $\phi^{\,\prime}(p)$  are both bound to vanish.  The integral
on the right hand side is non-singular  and in the limit $k\to p$ the 
integrand  goes to a finite limit $\Half\phi^{\,\prime}(p)/p+\frac{1}{8}\phi^{\,\prime\prime}(p)$. 
 This demonstrates explicitly that by using a subtraction technique 
it is perfectly possible to remove the singularity  converting the integral to a form
amenable for computation. Nevertheless, the subtraction scheme (\ref{I0}) would be insufficient
for solving an integral equation as it introduces 
unknown first  $\phi^{\,\prime}(p)$ and second derivative  $\phi^{\,\prime\prime}(p)$
 at the top of the unknown function. 
This also explains why the Nystrom method, 
which has been rather efficient in solving the Coulomb
bound state problem in momentum space \cite{lande}, does not work for the 
linear potential.   ( The calculation using Nystrom method
presented in  \cite{nys} is incorrect because  
 the infinite diagonal term in the potential matrix  
 has been simply omitted whereas  the  proposed  correction,
 given in their eq. (34),  is proportional to a logarithmically diverging integral.)
\par
In the early attempts to overcome this difficulty
 the singularity  was removed by hand, by introducing
 an arbitrary cut-off  \cite{chiu}\cite{eyre} in the potential.
 The resulting  non-singular  integral equation 
involving the modified  potential could be then solved
by standard methods. The unwelcome arifacts of the cutoff
might be eventually disposed of by perturbative methods \cite{eyre}.
However,  a more promising approach 
is to seek the wave function in the form of   
  an expansion in terms of a complete set of orthogonal
 basis functions.  The most common choice here has been the oscillator
or Sturmian basis  both of which have analytic Fourier-Bessel transforms
making them well suited in calculations where it is advantageous to work 
in configuration and momentum space simultaneously.  
 In a variational Ritz-type approach the  upper bounds of the
true eigenvalues could be computed by diagonalizing the corresponding 
Hamiltonian matrix (cf. \cite{var1}, \cite{var2},\cite{var3}).
 The expectation values of the energy can be
evaluated in momentum space and the potential expectation values in
configuration space.  
The expansion method could be used in a similar fashion to
solve the momentum space integral equation by means of the 
Galerkin method  \cite{spline}, \cite{maung}.  
With a judicious choice of the basis functions,  the singular 
integrals can be calculated analytically, or numerically.
Note, that in this case the integrand is a  known function and,
therefore, the subtraction technique, like the one outlined in (\ref{I0}),
is fully applicable.
%\par
 There are also non-variational approaches based on eigenfuction expansion
  such the collocation method \cite{spline},\cite{jean}, 
 or the  Multhopp \cite{wing}\cite{BB} technique.
 Keeping $N$ terms of the truncated expansion, the $N$ expansion coefficients can be
 determined from the requirement that the integral equation be exactly satisfied
 at $N$ distinct values of the momentum variable.    
The semi-spectral Chebyshev method developed in this paper
also belongs to the last group. However, the Chebyshev series, 
after reshuffling takes the form of an interpolative formula. 
In consequence, the expansion coefficients and
the function values taken at the mesh-points are connected by a linear
relation (cf. \cite{deloff}). Thus, put in a nut-shell, the underlying idea is to solve
the integral equation {\it exactly} on the Chebyshev mesh and, subsequently,
interpolate by means of a high degree polynomial.
 The plan of the presentation is as follows.
In the next section we set the necessary background 
deriving the hypersigular integral equation
associated with the Coulomb-plus-linear potential in momentum space.
Upon introducing the Chebyshev mesh and using the interpolative
formula for the wave 
function,  the integral equation is converted into
an algebraic eigenvalue problem. 
This is the ultimate form because the eigenvalue problem 
can be solved with the aid of standard library procedures.
 Section 3 is devoted to a numerical test where we compare 
the momentum space calculations with the results obtained by
solving the Schr\"{o}dinger equation in configuration space.
Finally in the last section we present our conclusions. 
%%%%%%%%%%%%%%%%%%%%%%%%%%%%%%%%%%%%%%%%%%%%%%%%%%%%%%%%%%%%%%%%%%%%
\section{Solution of the singular integral equation}
%%%%%%%%%%%%%%%%%%%%%%%%%%%%%%%%%%%%%%%%%%%%%%%%%%%%%%%%%%%%%%%%%%%%
The Coulomb-plus-linear potential considered in this paper is 
$V(r)= V^{(C)}(r)+V^{(L)}(r)$ with
\begin{equation}
 V^{(C)}(r)= -\alpha/r; \quad V^{(L)}(r)=r/a^2
 \label{prel1}
\end{equation}  
where the ''coupling'' $\alpha$ is dimensionless and the parameter
$a$ has a dimension of length ($\hbar=c=1$ units are adopted hereafter). 
Both parameters are assumed to be provided.
In momentum space the wave function $\phi_\ell(k)$ with 
orbital momentum $\ell$ obeys
the partial wave Schr\"{o}dinger equation 
 \begin{equation}
(E- k^2/2\mu)\;\phi_\ell(k)=
\int_0^\infty\,V_\ell(k,k')\,\phi_\ell(k')\,k^{\prime\, 2}\,\D k'
  \label{prel2}
 \end{equation}
 where $\mu$ is the quark-antiquark reduced mass, $E$ is the binding energy and
 $V_\ell(k,k')$ denotes the
 $\ell$-th partial wave projection of the local potential $V(r)$ 
  \begin{equation}
          V_\ell(k',k)= \dfrac{2}{\pi}
        \int_0^\infty j_\ell(k'r)\,V(r)\,j_\ell(kr)\,r^2\,\D r,
          \label{prel3}
  \end{equation}
where $j_\ell(x)$ is the spherical Bessel function \cite{abramowitz}.
Strictly speaking,  upon inserting  (\ref{prel1}) in (\ref{prel3}),
 we obtain a divergent integral but
a customary regularizing procedure to overcome this difficulty is first to 
 multiply $V(r)$ by a screening factor
$\E^{-\eta r}$ enforcing convergence
and then set $\eta\to 0$ in the result. Applying this procedure,
the Fourier transform  (\ref{prel3})
of a power--law potential $v(r)=r^{2n-1},\;n=0,1,2,...$  can be effected
in an analytic form \cite{BB}
\begin{equation}
   \lim_{\eta\to 0}\,\dfrac{2}{\pi}
     \int_0^\infty j_\ell(k'r)\,\E^{-\eta r}
        \,r^{2n+1}\,j_\ell(kr)\,\D r=
  \dfrac{(2n)!}{2^n\,n!\,\pi (k\,k')^{n+1}}\; Q_\ell^n (z)
          \label{prel4}
 \end{equation}
where $z=(k^2+k^{\prime\,2})/2kk'$ and the $Q_\ell^n(z)$ denotes $n$--th
derivative of the Legendre function of the second kind with respect
to the argument $z$ (formula (5) in \cite{BB} contains a misprint).
Setting $n=0$ and $n=1$ in (\ref{prel4}) we obtain, respectively, the kernels for
the Coulomb (C) and the linear  potential (L) 
\begin{equation}
 V^{(C)}_\ell(k,k')=-\alpha \, Q_\ell(z) /(\pi kk');
 \quad
 V^{(L)}_\ell(k,k')= Q_\ell^\prime(z)/[\pi (akk')^2].
\label{prel5}
\end{equation} 
The Coulomb part of the kernel exhibits a logarithmic singularity for  $k'=k$
contained in the Legendre function. Indeed,  the latter can be written as  
\begin{equation}
 Q_\ell(z)=P_\ell(z)\,Q_0(z) - w_{\ell-1}(z)
\label{prel6}
\end{equation}
where
\begin{equation}
 Q_0(z)=\Half\log|(1+z)/(1-z)|=\log|(k+k')/(k-k')|
\label{prel6a}
\end{equation}
with $ P_\ell(z)$ being a Legendre polynomial.
It is understood that
the last term in (\ref{prel6}) should be absent for $\ell=0$ 
whereas for $\ell>0$ it
 assumes the form of a polynomial in $z$ (cf. \cite{abramowitz})
given by the expression
\begin{equation}
        w_{\ell-1}(z)=
 \displaystyle\sum_{n=1}^\ell\frac{1}{n}P_{n-1}(z)\,P_{\ell-n}(z).
\label{prel7}
\end{equation}
The kernel associated with the linear potential given in (\ref{prel5}), 
in addition to the
logarithmic singularity, exhibits also a second order pole, as may be
seen by performing explicitly the differentiation in (\ref{prel6}) 
\begin{equation}
 Q_\ell^\prime(z)=P_\ell^\prime(z)\,Q_0(z)
 + P_\ell(z)\,Q_0^\prime(z)
 - w_{\ell-1}^\prime(z)
\label{prel8}
\end{equation}
with
\begin{equation}
 Q_0^\prime(z)=\dfrac{1}{1-z^2}
=-\left(\dfrac{2kk'}{k'+k}\right)^2\dfrac{1}{(k'-k)^2}.
\label{prel8a}
\end{equation}
The second order pole  given by (\ref{prel8a})
can be eliminated from the integral equation (\ref{prel2})  and to this end
integration by parts is applied to  this  term.
Quite generally,  this procedure  gives 
\begin{equation}
\int_0^\infty \dfrac{f(k,k')\,\phi_\ell(k')\,\D k'}{(k'-k)^2}=
\int_0^\infty \dfrac{\D k'}{k'-k}\,  \dfrac{\partial }{\partial k'}\, \left[f(k,k')\,\phi_\ell(k')\right]
\label{prel9}
\end{equation} 
where the unspecified  function  $f(k,k')$ needs to be integrable.
The above formula holds because the wave function $\phi_\ell(k')$
vanishes when  $k'$ tends to either of  the integration end points.  
 The resulting Cauchy principal value  integral in   (\ref{prel9}) can be computed 
by using the dedicated Chebyshev quadrature given in \cite{deloff}. 
Nevertheless,  the  lowering of the order of the pole outlined above has its price and in the integral on the
right hand side  of (\ref{prel9})  the derivative of the unknown wave function will appear.
As we shall see in a moment,  the semi-spectral Chebyshev method is well suited
to handle such situation.          
\par
It will be convenient for us using $1/a$ as the unit of energy,  passing to dimensionless
quantities: $ \epsilon\equiv Ea,\;x\equiv ka,\;x' \equiv k'a$.  The resulting integral equation 
\begin{equation}
\label{prel10}
\begin{split}
\left(\epsilon -\dfrac{x^2}{2\mu a} \right)\phi_\ell(x)=
\dfrac{1}{\pi x^2}\int_0^\infty\left\{P_\ell^{\,\prime} (z) \, \log\left|\dfrac{x'+x}{x'-x}\right|
-w_{\ell-1}^{\,\prime}(z)\right\}\phi_\ell(x')\,\D x'
\hspace*{.2\textwidth}
\\ \mbox{}-
\dfrac{4}{\pi}\int_0^\infty
\dfrac{\D x'}{x'-x}\;\left\{\chi_\ell(x')+\phi_\ell(x')\,\dfrac{\partial}{\partial x'} \right\}
 \dfrac{x^{\prime\, 2} P_\ell(z)}{(x'+x)^2}-
\hspace*{.2\textwidth}
\\ \mbox{}-
 \dfrac{\alpha}{\pi x}\int_0^\infty\left\{P_\ell(z)\, \log\left|\dfrac{x'+x}{x'-x}\right|
-w_{\ell-1}(z)\right\}\phi_\ell(x')x'\,\D x' 
\hspace*{.2\textwidth}
\end{split}
\end{equation} 
  involves  two dimensionless parameters:  $\alpha$ and $2\mu a$.  Prime on a function
of $z$ denotes in (\ref{prel10}) the derivative with respect to the argument.
The derivative of the wave function appearing in the integrand of the second integral   
in (\ref{prel10})
has been regarded as an additional function  $\chi_\ell(x)$ to be determined.
In order to complete our scheme  the integral  equation (\ref{prel10})
needs to be supplemented with a complementary equation
\begin{equation}
\D \phi_\ell(x)/\D x=\chi_\ell(x)
\label{prel11}
\end{equation}
and we end up with two equations for two unknown functions: $\phi_\ell(x)$ and $\chi_\ell(x)$.
\par
The system (\ref{prel10})--(\ref{prel11}) is amenable for computation and 
the integral equation will be turned into a finite matrix equation.  
 As a preliminary step,  the semi-infinite domain of the independent variable $x$
will be  mapped onto a finite interval $(-1,1)$.  
Among endless possibilities  perhaps the simplest is the rational mapping
\begin{equation}
\label{prel12}
x=\sigma (1+t)/(1-t),
\end{equation}   
where $t\in (-1,1)$ and $\sigma$ is a numeric parameter 
at our disposal  providing additional control of
 the rate of convergence.  We tried some other mappings, specifically trigonometric
($x=\sigma \tan[(\pi/4)(1+t)]$), or logarithmic ($x=\sigma \log[(3+t)/(1-t)]$) but they 
did not bring noticible  improvement in the problem under consideration.
The semi-spectral Chebyshev method uses Chebyshev polynomials 
 as the basis functions. The Chebyshev polynomial of the first kind $T_N(t)$ of the order
$N$ is defined by the formula  
\begin{equation}
T_N(t)=\cos[N\, \arccos(t)]
\label{prel15}
\end{equation}
and has $N$ zeros in the interval $(-1,1)$, located at the points 
\begin{equation}
\label{prel13}
t_i = \cos[\pi(i-\Half)/N];   \quad    i=1,2,...,N.
\end{equation}
 In the following the variable $t$ will be discretized by  using  the classical  Chebyshev
mesh (\ref{prel13})  in which case
 $N$  becomes the order of approximation to be selected by the user.  
 The  semi-spectral
Chebyshev method interpolates the unknown function $f(t)$ on the Chebyshev mesh (\ref{prel13})
\begin{equation}
f(t)=\sum_{i=1}^N f(t_i)\,G_i (t),  
\label{prel14}
\end{equation}
where $G_i(t)$ denotes the cardinal function with the property $G_i(t_j)=\delta_{ij}$. These functions
can be constructed as superpositions of Chebyshev polynomials
\begin{equation}
  G_j(t)=\dfrac{2}{N}\sum_{i=1}^N {\rm '\mbox{} } \ T_{i-1}(t_j)\,T_{i-1}(t),
\label{prel16}
\end{equation}
where the  
 primed sigma denotes a summation in which the first term should be halved.
By taking advantage of the interpolative formula (\ref{prel14}),
 the differentiation or integration of a function reduces to differentiation or
integration of Chebyshev polynomials which in most cases is elementary and can be performed  in
an analytic form.  In consequence, the array  containing the values 
of  the derivative computed at the grid-points will be
connected to similar array representing  the function by a linear transformation
\begin{equation}
\left\{\dfrac{\D f(t)}{\D t}\right\}_{t=t_i}=\sum_{j=1}^N D_{ij} \, f(t_j),\quad i=1,2,...,N
\label{prel17}
\end{equation}
where $D_{ij}$ is easily computed  numerical matrix (cf. \cite{deloff}).        
There are also various  integration rules available. 
 Assuming  that the function $f(t)$ is non-singular in the integration domain, we have 
\begin{equation}
\int_{-1}^1 f(t) \,\D t =\displaystyle \sum_{i=1}^N w_i\, f(t_i),
\label{prel18}
\end{equation}
which is Gauss-Chebyshev integration in which the weighting function is equal
to unity. The weights $w_i$ are all positive and their sum equals to 2. Similar rules can be
derived for singular integrals.
The Cauchy principal value integration can be performed using the
automated quadrature rule
  \begin{equation}
\int_{-1}^1\dfrac{ f(t) \,\D t}{t-\tau} = \displaystyle\sum_{i=1}^N \omega_i(\tau)\, f(t_i),
\label{prel19}
\end{equation}
where it is assumed that $\tau\in<-1,1>$.  When $\tau$ coincides with either of the 
integration end-points the integral is undefined. The dedicated weighting functions $\omega_i(\tau)$
can be calculated analytically  and  exhibit logarithmic end-point singularity for $\tau=\pm 1$.
Similar rule can be obtained for a weakly singular integral
   \begin{equation}
\int_{-1}^1 f(t) \,\log|t-\tau|\,\D t = \displaystyle\sum_{i=1}^N \Omega_i(\tau)\, f(t_i),
\label{prel20}
\end{equation}
where it is assumed that $\tau\in(-1,1)$.  
In contrast with the previous case,  $\log|t-\tau|$ singularity is integrable
and   the dedicated weighting functions $\Omega_i(\tau)$
do exist even   when $\tau$ coincides with either of the 
integration end-points. For explicit analytic expressions for  all of the weighting
functions introduced above the reader is referred to  \cite{deloff}.
\par
To arrive at the ultimate finite matrix eigenvalue problem, as the first step,  we  map 
both,  the external ($x$),  and the internal ($x'$)
variable  onto the $(-1,1)$ interval with the aid of (\ref{prel12}).  Subsequently, the problem
 is discretized by putting the external variable on the Chebyshev mesh  (\ref{prel13}), 
at the same time replacing all integrations in  (\ref{prel10}) by summations, 
 following the appropriate Chebyshev  rules listed  above.  In practice this
procedure leads to  a chain of  substitutions  to be made in 
the integrals occurring in  (\ref{prel10}),  {\it viz.}
\begin{equation*}
 x\to x_i =\sigma(1+t_i)/(1-t_i);\quad     \phi_\ell(x)\to\phi_\ell(x_i)\equiv X_i;
\end{equation*}
and
\begin{equation*}
 x'\to x_j =\sigma(1+t_j)/(1-t_j);\quad     \phi_\ell(x')\to\phi_\ell(x_j)\equiv X_j; 
\end{equation*}
where  $X_i$ are the unknown mesh values of the wave function to be determined. 
The derivative  $\chi_\ell(x_j)$ is eliminated in favor of $X_j$ with the aid
of the $D_{ij}$ matrix, accounting for the change of variables
\begin{equation*}
 \chi_\ell(x_j) = \dfrac{(1-t_j)^2}{2\sigma}\sum_{k=1}^N D_{jk}\,X_k. 
\end{equation*}
Further substitutions associated with integration,  respectively, are   
\begin{equation*}
  \D x'                           \to  2\sigma w_j/(1-t_j)^2,  
\end{equation*}
for non-singular integrals
\begin{equation*}
  \dfrac{\D x' }{x' - x}                \to  \omega_j   (t_i)\,   \dfrac{1 - t_i }{1 - t_j },
\end{equation*}
for principal value integral, and
\begin{equation*}
  \log \left| \dfrac{x'+x}{x'-x} \right|\, \D x'     \to  2 \sigma \,\dfrac{w_j\,\log|1-t_i\,t_j| - \Omega_j (t_i)}{(1-t_j)^2}
\end{equation*}
for integrals involving logarithmic singularity.   Finally, all integrations will be effected by 
carrying out a summation over  $j$.  It is worth noting that the diagonal terms $i=j$ 
are always finite and all singularities are under control.
\par
When the  indicated above  manipulations have been accomplished, we end up with a  homogeneous
system of  $N$  algebraic equations in which the $N$ unknowns are the 
  mesh-point values of the wave function ($X_j$) and the Schr\"{o}dinger equation
takes the desired finite matrix form 
\begin{equation}
 \sum_{j=1}^N \left(
  V_{ij}+\dfrac{x_i^2}{2\mu a}\, \delta_{ij} -\epsilon \right)\,X_j = 0.
\label{prel21}
\end{equation} 
The non-symmetric matrix $V_{ij}$  represents here  the potential and results
from evaluating the  integrals  occurring on the
right hand side of (\ref{prel10}) (the explicit form of $V_{ij}$ is rather lengthy and will
not be quoted here).
When the kinetic energy term is lumped together with $V_{ij}$ into a single matrix,  eq. (\ref{prel21}) presents  a standard algebraic eigenvalue problem. 
If need arises, the non-relativistic Schr\"{o}dinger
equation (\ref{prel21}) can be easily converted to the relativistic form (\ref{I1})
in the center-of-mass frame by changing just the kinetic energy term
\begin{displaymath}
 x_i^2/(2\mu a)\to \sqrt{x_i^2+(a\,m_1)^2}+\sqrt{x_i^2+(a\,m_2)^2}-a\,(m_1+m_2).
\end{displaymath}
 Our calculational scheme is now complete and for assigned values of $\ell$ and
 two dimensionless parameters
 $s\equiv 1/2\mu a$ and $\alpha$ specifying the strength of the two potentials 
 in (\ref{prel1}), we are in the position to determine numerically 
 the value of the binding energy $\epsilon(\ell,s,\alpha)$.
 In the particular case $\ell=0$ and $\alpha=0$
 the exact result is known and the binding energy is 
 $\epsilon(0,s,0)=-s^{2/3}\, z_\nu$  where $z_\nu$ with $\nu=1,2,3,...$
 denotes a zero of the Airy function $Ai(z)$ (cf. \cite{abramowitz}).  
%%%%%%%%%%%%%%%%%%%%%%%%%%%%%%%%%%%%%%%%%%%%%%%%
\section{Numerical test}
%%%%%%%%%%%%%%%%%%%%%%%%%%%%%%%%%%%%%%%%%%%%%%%
 We start the numerical test with the Coulomb bound state problem 
 leaving out the first two integrals on the right hand side of  (\ref{prel10}).
 The hydrogen-like bound state problem in momentum space
has already been considered in \cite{deloff} but to
determine the bound states we solved the secular equation.
It is therefore of interest to repeat the Coulomb bound-state
calculation in which the energy spectrum is obtained by 
solving the algebraic eigenvalue problem  (\ref{prel21}).
The latter procedure is much simpler as there is no need to
solve a transcendental equation. 
In all our computations 
we were using the linear algebra package LAPACK \cite{lapack}
as our eigenvalue solver. The results for the Coulomb
potential are displayed in Table \ref{table1}. Since in this case the 
exact eigenenergies are known analytically we present
 the absolute value of the relative error
on each level as a function of the mesh size $N$.
The nodal quantum number $n$ enumerates the
 the different bound states  for a fixed $\ell$
 with $n=0$ corresponding to the ground state.  
 We wish to recall that 
with non-symmetric matrices the accuracy of the standard library procedures
is believed to be not as good as in the case of symmetric matrices.
Nevertheless, as seen from Table \ref{table1}, the convergence rate
is exponential and  $N=80$  is sufficient for securing machine accuracy.
There are not very many methods available that would be capable of
achieving such a high precision.
For comparison, in the last raw (entries in parenthesis) 
we give the relative error corresponding
to the traditional method using  the
 subtraction scheme \cite{lande} in which case
the resulting eigenvalue problem is symmetric.
The advantage of the semi-spectral method is manifest.
\par
As our next test we take on the
linear potential setting $\alpha=0$ in (\ref{prel21})
 and  putting for simplicity $s=1$ in our computations.
The resulting binding energies  $\epsilon$ 
for different $\ell$ values are displayed in
Table \ref{table2} using the same conventions as in Table \ref{table1}.
 For $\ell=0$, as the exact values we take
the zeros of the Airy function tabulated in \cite{abramowitz}.
For $\ell>0$ the values marked as exact have been computed by solving the appropriate
Schr\"{o}dinger equation in configuration space. For this purpose we used
the ingenious algorithm developed in \cite{wien}. The code from \cite{wien}
 has been revamped for obsolescent features and
 the original Runge-Kutta driver
advancing the solution from $x$ to $x+h$ has been replaced by a more accurate
driver based on Chebyshev approximation as 
described in \cite{deloff}. After the above changes, 
the typical relative error in all
considered here cases was estimated to be of the order of $10^{-11}$.
As a cross-check, we succeeded in
reproducing the exact results for $\ell=0$ up to eleven significant digits. 
To obtain the entries in table \ref{table2}
for each $\ell$ value we needed to solve the algebraic 
eigenvalue problem (\ref{prel21}) and
 in nearly all considered here cases 
we managed to get seven significant figures which is more than adequate 
in all practical applications. Our results have been obtained keeping quite moderate
approximation order $N=100$. Only the $\ell=0$ case which  was more stubborn
forced us to go to larger $N$. 
It is apparent from table \ref{table2} that the solutions
are very stable with respect to increasing $N$ albeit the rate of convergence is 
no longer exponential. In fact, it is quite slow when compared with
the Coulomb case.
 Making such comparison, however, it has to be kept in mind
 that in the linear potential case
we need to determine  two unknown functions  
(wave function and its derivative) rather than one 
and therefore $N$ should have  been doubled if we wanted 
to keep the same number of points per function. 
 Other than that, there is probably a good deal of cancellation
across the pole and this might be responsible for some loss of accuracy. 
\par
Finally, we are going to consider the case where both, the Coulomb and the linear
potential are present.  The quark-antiquark potential has been adopted from 
a realistic study \cite{quigg} of charmonium $(c\bar{c})$ and bottomium $(b\bar{b})$
$V(r)=-\alpha/r+\beta\,r$ where we stick to the parameter values
provided in \cite{quigg}, namely 
\begin{equation}
 \alpha=0.50667,\quad \beta=0.1694\;GeV^2,\quad m_c=1.37\;GeV,\quad m_b=4.79\;GeV. 
\label{q}
\end{equation}
The results of our computations are presented in Table \ref{table3}. The  
quarkonium masses $M$ displayed there 
 have been obtained from the expression $M=2m_q+E$ where $m_q$ is the
quark(antiquark) mass. To determine the binding energy $E$ the appropriate
 non-relativistic Schr\"{o}dinger equation was solved in both, the momentum 
and the configuration space.  
 As seen from Table \ref{table3} there is excellent
agreement between these two approaches. 
%%%%%%%%%%%%%%%%%%%%%%%%%%%%%%%%%%%%%%%%%%%
\section{Summary and Conclusion}
%%%%%%%%%%%%%%%%%%%%%%%%%%%%%%%%%%%%%%%%%%%
The aim of this paper was to demonstrate the strength of
the semi-spectral Chebyshev method in solving integral 
equations whose kernels exhibit singularities of 
the Cauchy or the logarithmic type. 
Such equations may be encountered
 in quantum mechanics as has been exemplified 
by considering the Coulomb-plus-linear potential bound state problem
in momentum space. The latter problem is considered in this
work for illustrative purposes and therefore we have gone in some details.
The semi-spectral Chebyshev method has many advantageous features.
First, it is very easy to use 
since it is based on a polynomial
interpolation where both, the mesh and the polynomials, can be readily 
obtained in an analytic form.
 Second, the programming is exceedingly simple because differentiation or integration
of polynomials can be performed analytically and on a mesh these operations
take the form of matrix multiplications.  The presented method is well suited
to handle singular integral equations (with Cauchy or logarithmic singularities)
because automatic quadratures are provided for evaluating singular integrals.
 This allows for a quick and seamless discretization and
 since the integrals involving singular kernels have finite diagonal 
 elements the Nystrom method is still applicable.
 Ultimately, the integral equation
is converted into an algebraic eigenvalue problem which can be solved directly
 by standard library procedures.
There is no need to solve a complicated transcendental
equation. Third, the method is highly accurate. This is because the approximation
is global basing on a polynomial of a high degree.
The eigenvectors contain the wave function values on the mesh and can be used
to calculate various expectation values. If this is not enough,
 once the integral equation
has been solved exactly on the mesh, the solution at an arbitrary point
may be immediately obtained by interpolation. 
In conclusion, with the aid of the semi-spectral Chebyshev method  
the solution of a singular integral equation  becomes
no more difficult than the solution of a Fredholm equation.

%%%%%%%%%%%%%%%%%%%%%%%%%%%%%%%%%%%%%%%%%%%%%%%

%%%%%%%%%%%%%%%%%%%%%%%%%%%%%%%%%%%%%%%%%%%%%%%%%%%%%%%%%%%%%%%%%%%%%%%%%%%%%%%%%

%%%%%%%%%%%%%%%%%%%%%%%%%%%%%%%%%%%%%%%%%%%%%%%
\newpage
%%%%%%%%%%%%%%%%%%%%%%%%%%%%%%%%%%%%%%%%%%%%%%%%%%%%%%%%%%%%%%%%%%%%%%%%%%%%%%
 \begin{table}[ht]
 %\label{table1}
\caption{Relative errors on the computed Coulomb binding energies.
The corresponding errors appropriate to traditional method based on subtraction
are given in parenthesis. }
\begin{center}
\vspace{2mm}
\begin{tabular}{rlllll}
      &            &            &   $\ell=0$         &            &          \\
 $N\;$&$\; n=0$&$\; n=1$&$\; n=2$&$\; n=3$&$\; n=4$\\ \hline 
$40$&$4\times 10^{-12}$&$3\times 10^{-10}$&$4\times 10^{ -9}$&$3\times 10^{ -8}$&$2\times 10^{ -7}$\\
$60$&$2\times 10^{-13}$&$1\times 10^{-11}$&$2\times 10^{-10}$&$1\times 10^{ -9}$&$6\times 10^{ -9}$\\
$80$&$2\times 10^{-14}$&$1\times 10^{-12}$&$2\times 10^{-11}$&$1\times 10^{-10}$&$6\times 10^{-10}$\\
$80$&$(4\times 10^{ -5})$&$(9\times 10^{ -5})$&$(2\times 10^{ -4})$&$(4\times 10^{ -4})$&$(6\times 10^{ -4})$\\      \hline\hline 
                      %  &&&&&\\ 
       &            &            &   $\ell=1$         &            &          \\
 $N\;$&$\; n=0$&$\; n=1$&$\; n=2$&$\; n=3$&$\; n=4$\\ \hline 
$40$&$8\times 10^{-13}$&$2\times 10^{-13}$&$3\times 10^{ -9}$&$6\times 10^{ -8}$&$5\times 10^{ -7}$\\
$60$&$2\times 10^{-14}$&$4\times 10^{-13}$&$3\times 10^{-12}$&$1\times 10^{-11}$&$2\times 10^{-10}$\\
$80$&$2\times 10^{-15}$&$4\times 10^{-14}$&$3\times 10^{-13}$&$1\times 10^{-12}$&$2\times 10^{-12}$\\
$80$&$(7\times 10^{ -6})$&$(5\times 10^{ -5})$&$(2\times 10^{ -4})$&$(5\times 10^{ -4})$&$(1\times 10^{ -3})$\\      \hline\hline 
                          % &&&&&\\ 
       &            &            &   $\ell=2$         &            &          \\ 
 $N\;$&$\; n=0$&$\; n=1$&$\; n=2$&$\; n=3$&$\; n=4$\\ \hline  
$40$&$3\times 10^{-12}$&$2\times 10^{-10}$&$2\times 10^{ -7}$&$5\times 10^{ -6}$&$8\times 10^{ -5}$\\
$60$&$3\times 10^{-15}$&$6\times 10^{-14}$&$1\times 10^{-12}$&$7\times 10^{-10}$&$2\times 10^{ -8}$\\
$80$&$2\times 10^{-16}$&$2\times 10^{-15}$&$4\times 10^{-14}$&$4\times 10^{-13}$&$6\times 10^{-12}$\\
$80$&$(1\times 10^{-5})$&$(6\times 10^{-5})$&$(3\times 10^{-4})$&$(7\times 10^{-4})$&$(2\times 10^{-3})$\\      \hline\hline
                     % &&&&&\\
      &            &            &   $\ell=3$         &            &          \\
 $N\;$&$\; n=0$&$\; n=1$&$\; n=2$&$\; n=3$&$\; n=4$\\[1mm] \hline 
$40$&$2\times 10^{ -9}$&$2\times 10^{ -7}$&$9\times 10^{ -6}$&$3\times 10^{ -4}$&$3\times 10^{ -3}$\\
$60$&$1\times 10^{-12}$&$3\times 10^{-12}$&$8\times 10^{-11}$&$2\times 10^{ -8}$&$6\times 10^{ -7}$\\
$80$&$1\times 10^{-13}$&$2\times 10^{-12}$&$6\times 10^{-12}$&$1\times 10^{-10}$&$5\times 10^{-10}$\\
$80$&$(8\times 10^{-5})$&$(3\times 10^{-5})$&$(3\times 10^{-4})$&$(1\times 10^{-3})$&$(1\times 10^{-3})$\\      \hline\hline

\end{tabular}
\end{center}
\label{table1}
\end{table} 

%%%%%%%%%%%%%%%%%%%%%%%%%%%%%%%%%%%%%%%%%%%%%%%%%%%%%%%%%%%%%%%%%%%%%%%%%%%%%%%
\newpage
%%%%%%%%%%%%%%%%%%%%%%%%%%%%%%%%%%%%%%%%%%%%%%%%%%%%%%%%%%%%%%%%%%%%%%%%%%%%%%%
 \begin{table}[ht]
\caption{Binding energy $\epsilon $ for a linear potential.}
\begin{center}
\vspace{2mm}
\begin{tabular}{rlllll}
      &            &            &   $\ell=0$         &            &          \\

 $N\;$ &  $\quad n=0$  &  $\quad n=1$  &   $\quad n=2$    &  $\quad n=3$     &   $\quad n=4$  \\
\hline
%      &            &            &            &            &          \\
  50  &  2.338034  &  4.087928  &  5.520416  &  6.786654  &  7.943940\\
 100  &  2.338099  &  4.087947  &  5.520543  &  6.786702  &  7.944111\\
 150  &  2.338105  &  4.087949  &  5.520555  &  6.786706  &  7.944127\\
 200  &  2.338106  &  4.087949  &  5.520558  &  6.786707  &  7.944131\\
 250  &  2.338107  &  4.087949  &  5.520559  &  6.786708  &  7.944132\\
 300  &  2.338107  &  4.087949  &  5.520559  &  6.786708  &  7.944133\\
{\it exact} &  2.338107  &  4.087949  &  5.520560  &  6.786708  &  7.944134\\
\hline\hline
      &            &            & $\ell=1$           &            &          \\

 $N\;$  &  $\quad n=0$ &  $\quad n=1$   &   $\quad n=2$    &  $\quad n=3$     &   $\quad n=4$  \\
\hline
 %     &            &            &            &            &          \\
  50  &  3.361254  &  4.884452  &  6.207617  &  7.405649  &  8.515212\\
 100  &  3.361255  &  4.884452  &  6.207623  &  7.405665  &  8.515234\\
{\it exact} &  3.361254  &  4.884452  &  6.207623  &  7.405665  &  8.515234\\
\hline\hline
      &            &            &  $\ell=2 $          &            &          \\

   $N\;$  &  $\quad n=0$ &  $\quad n=1$  &   $\quad n=2$    &  $\quad n=3$     &   $\quad n=4$  \\
\hline
 %     &            &            &            &            &          \\
  50  &  4.248183  &  5.629693  &  6.868774  &  8.009828  &  9.075383\\
 100  &  4.248182  &  5.629708  &  6.868883  &  8.009703  &  9.077003\\
{\it exact} &  4.248182  &  5.629708  &  6.868883  &  8.009703  &  9.077003\\
\hline\hline
      &            &            &  $\ell=3$          &            &          \\

$N\;$  &  $\quad n=0$   &  $\quad n=1$ & $\quad n=2$ &  $\quad n=3$  &   $\quad n=4$  \\
\hline
  %    &            &            &            &            &          \\
  50  &  5.050918  &  6.331874  &  7.504206  &  8.593338  &  9.632163\\
  80  &  5.050926  &  6.332115  &  7.504646  &  8.597127  &  9.627263\\
{\it exact} &  5.050926  &  6.332115  &  7.504646  &  8.597117  &  9.627267\\
\hline\hline
\end{tabular}
\end{center}
\label{table2}
\end{table}

%%%%%%%%%%%%%%%%%%%%%%%%%%%%%%%%%%%%%%%%%%%%%%%%%%%%%%%%%%%%
\newpage
%%%%%%%%%%%%%%%%%%%%%%%%%%%%%%%%%%%%%%%%%%%%%%%%%%%%%%%%%%%
 \begin{table}[ht]
% \label{table3}
\caption{
Charmonium $(c\bar{c})$ and bottomium $(b\bar{b})$ masses 
(all entries in GeV) computed from
the Coulomb-plus-linear potential \cite{quigg} $V(r)=-\alpha/r+\beta r$ with 
the parameters given in (\ref{q}). The upper (lower) values 
result from a calculation conducted in momentum (configuration) space using
non-relativistic Schr\"{o}dinger equation. In all momentum space computations 
the mesh size was $N=80$. }
\begin{center}
\vspace{2mm}
\begin{tabular}{lrcr}
\hline  
\hline  
             &           &  $m_c$ = 1.37 & \\
\hline  
             &   $n=0$   &   $n=1$   &   $n=2\;$ \\ 
\hline
 $\ell = 0$  &   3.0869  &   3.6748  &   4.1094\\
             &   3.0869  &   3.6748  &   4.1093\\   
\hline
 $\ell = 1$  &   3.4988  &   3.9544  &   4.3388\\
             &   3.4987  &   3.9543  &   4.3388\\
\hline
 $\ell = 2$  &   3.7868  &   4.1868  &   4.5407\\
             &   3.7868  &   4.1868  &   4.5407\\
\hline\hline 
             &           &  $m_b$ = 4.79 & \\
\hline  
             &   $n=0$   &   $n=1$   &   $n=2\;$ \\ 
 \hline
 $\ell = 0$  &   9.4550  &  10.0105  &  10.3423\\
             &   9.4547  &  10.0104  &  10.3422\\
\hline
 $\ell = 1$  &   9.9171  &  10.2582  &  10.5318\\
             &   9.9170  &  10.2581  &  10.5318\\
\hline
 $\ell = 2$  &  10.1555  &  10.4385  &  10.6838\\
             &  10.1554  &  10.4385  &  10.6410\\
\hline\hline

\end{tabular}
\end{center}
\label{table3}
\end{table} 

\end{document}